\documentclass[review]{elsarticle}
\usepackage{float}
\usepackage{geometry}
\usepackage{setspace}

\usepackage{epstopdf}
\usepackage{lineno,hyperref}
\usepackage{amsmath}
\usepackage{makecell,multirow,diagbox}
\journal{Journal of \LaTeX\ Templates}

\begin{document}

\begin{frontmatter}

\title{Differential symbolic entropy in nonlinear dynamics complexity analysis}
\author[mymainaddress]{Wenpo Yao}

\author[mysecondaryaddress]{Jun Wang\corref{mycorrespondingauthor}}
\cortext[mycorrespondingauthor]{Corresponding author}
\ead{wangj@njupt.edu.cn}

\address[mymainaddress]{School of Telecommunications and Information Engineering, Nanjing University of Posts and Telecommunications, Nanjing 210003, China}
\address[mysecondaryaddress]{School of Geography and Biological Information, Nanjing University of Posts and Telecommunications, Nanjing 210023, China}

\begin{abstract}
Differential symbolic entropy, a measure for nonlinear dynamics complexity, is proposed in our contribution. With flexible controlling parameter, the chaotic deterministic measure takes advantage of local nonlinear dynamical information among three adjacent elements to extract nonlinear complexity. In nonlinear complexity detections of chaotic logistic series, DSEn (differential symbolic entropy) has satisfied complexity extractions with the changes of chaotic features of logistic map. In nonlinear analysis of real-world physiological heart signals, three kinds of heart rates are significantly distinguished by DSEn in statistics, healthy young subjects $>$ healthy elderly people $>$ CHF (congestive heart failure) patients, highlighting the complex-losing theory of aging and heart diseases in cardiac nonlinearity. Moreover, DSEn does not have high demand on data length and can extract nonlinear complexity at short data sets; therefore, it is an efficient parameter to characterize nonlinear dynamic complexity.
\end{abstract}


\end{frontmatter}

%\linenumbers

\section{Introduction}
The paradigm of deterministic chaos for nonlinear complex systems becomes increasingly popular for its attractive concept and successful applications in chaotic models and real-world complex systems \cite{Schreiber2000}. Nonlinear complexity measures, including fractal dimensions, correlation dimension, and Lyapunov exponent etc., are defined for chaotic dynamical systems and applied in physics, biology, meteorology, chemistry and so on \cite{Storgatz2000}. Some entropy methods, such as K-S entropy, approximate entropy \cite{Pincus1991}, sample entropy \cite{Richman2000,Keller2009,Chen2009}, permutation entropy \cite{Bandt2002,Bandt2016}, multiscale entropy \cite{costa2002MP,Costa2005MPE} and so forth \cite{Li2015,Isler2007,Yao2014} are also developed for these nonlinear dynamics. Among these nonlinear complexity measures, symbolic dynamics analysis, with basic idea of simplicity and efficiency, provides a rigorous way of observing dynamics with finite precision \cite{Bandt2002,Li2006,Hao1991}.

Symbolic time series analysis involves in symbolic transformation and measurements for these symbolic sequences \cite{Daw2003}. It reduces high requirements for data by transforming raw series into a finite number of states and mapping each state onto symbol from a given alphabet. For example, symbolic transfer entropy \cite{Staniek2008,Yao2017P} and multiscale symbolic entropy analysis \cite{Lo2015} improves original methods, like reducing high demands on data or sensitivity to noise, by taking advantage of symbolic dynamic analysis. Symbolizations of these methods can be grouped into the static range-partitioning and dynamic differenced-based approaches \cite{Daw2003,Yao2017D}, among which dynamical methods have high real-time features and are relatively insensitive to extreme noise spikes \cite{Daw1998,Finney1998}. Measures of symbolic sequence include direct visual histograms and quantitative measures based on classical statistics or information theory \cite{Daw2003}. The combinations of symbolization and entropy measures, belonging to information methods, play important roles in complexity detections and nonlinear dynamics analysis for their characteristics of simplicity, fast, insensitivity to noise etc. \cite{Li2006,Zanin2012,Wessel2000,Wessel2007}.

DSEn (differential symbolic entropy), targeting on informational properties of dynamical symbolic sequence, is proposed in our contributions. It extracts local nonlinear dynamic information from three adjacent elements and uses adjustable controlling parameter to improve flexibility in nonlinear complexity detections. Chaotic model, logistic map, and three groups of real-world physiological heart signals are applied to test nonlinear dynamic complexity detections of DSEn.

\section{Differential symbolic entropy}

\subsection{JK symbolization}
Symbolization plays important role in symbolic dynamic analysis. The symbolic procedure inevitably leads to the loss of part of statistical information; however, it simplifies time series analysis and contributes to dynamic complexity detection by extracting symbolic dynamic information \cite{Wessel2000,Wessel2007}.

A symbolization in works of J. Kurths et al. \cite{Kurths1995}, using typical local dynamic symbolization, conducts symbolic transformation by comparing relationships between adjacent symbols. Given time series $X=\{x_{1},x_{2},\cdots,x_{i},\cdots \}$, JK symbolization, being described as especially reflects dynamical properties of the record \cite{Kurths1995}, transforms time series into symbol sequence as Eq.~(\ref{eq1}).
\begin{eqnarray}
\label{eq1}
s_{i}(x_{i})=
  \left\{
       \begin{array}{lr}
          0: \Delta x > 1.5 \sigma_{\Delta} \\
          1: \Delta x > 0 \ and \ \Delta x \leq 1.5 \sigma_{\Delta} \\
          2: \Delta x > - 1.5 \sigma_{\Delta}  \ and \ \Delta x \leq  0 \\
          3: \Delta x \leq - 1.5 \sigma_{\Delta}
       \end{array}
  \right.
\end{eqnarray}
where $\Delta x = x_{i+1} - x_{i}$, and $\sigma_{\Delta}$ is variance of the adjacent measurement values.

JK symbolic transformation refines differences between neighboring elements, but it only considers two adjacent values and lack flexibility due to the fixed $1.5 \sigma_{\Delta}$.

\subsection{Differential symbolization}
Taking relationships of three consecutive elements into account, we propose differential symbolic transformation with flexible controlling parameter.

Considering time series $X=\{x_{1},x_{2},\cdots,x_{i},\cdots \}$, the differences between current element and its forward and backward ones are expressed as $D_{1} = \|x(i)-x(i-1)\|$ and $D_{2} = \|x(i)-x(i+1)\|$. Difference between $D_{1}$  and $D_{2}$ is calculated as $diff=D_{1}-D_{2}$ and $var$ is defined as $ \sqrt{(D_{1}^{2}+D_{2}^{2})/2}$ . The four-symbols differential symbolization with controlling parameter $\alpha$ is obtained through Eq.~(\ref{eq2}).

\begin{eqnarray}
\label{eq2}
s_{i}(x_{i})=
  \left\{
       \begin{array}{lr}
          0: diff \geq \alpha \cdot var \\
          1: 0 \leq diff < \alpha \cdot var  \\
          2: - \alpha \cdot var < diff < 0 \\
          3: diff \geq - \alpha \cdot var
       \end{array}
  \right.
\end{eqnarray}

The symbolization in Eq.~(\ref{eq2}) takes advantage of more detailed local information to complexity measures than JK symbolic transformation.

\begin{figure}[htb]
  \centering
    \includegraphics[width=11cm,height=4cm]{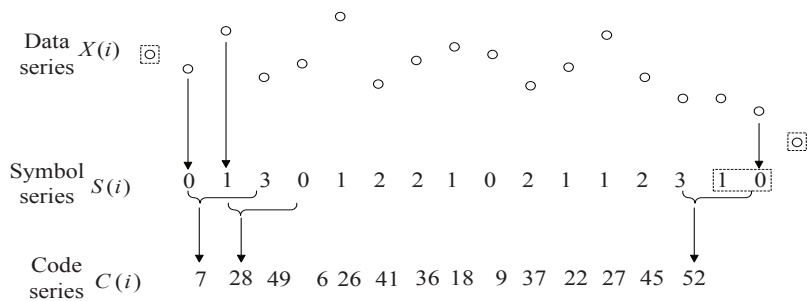}
  \caption{Process of symbolization and coding (data and symbols in virtual frame will not be symbolized or encoded). In creation of code series, symbol length $m$ is 3 and delay $\tau$ is 1. The first and last elements will not be transformed according to the determination of symbolization, and the last n-1 bit symbols are not encoded for this encoding process}
  \label{fig1}
\end{figure}

Construction of symbol sequences, or words, is the next step by collecting groups of symbols together in temporal order. This coding process is to create symbols templates or words with finite symbols and has some similarities to embedding theory for phase space construction. Symbol sequence will be coded into m-bit series $C(i)$ and there are $4^{m}$ symbols in coded series considering the 4-symbols differential symbolization. Taking 3-bit coding as example, code for 'abc' can be $c(i)=a*n^{2}+b*n+c$ where 'n' should not be smaller than the number of symbols' types. And a procedure of symbolization and coding is illustrated in Fig.~\ref{fig1}. The probability of each code symbol is $ P(\pi)=[p(\pi_{1}),p(\pi_{2}),\ldots,p(\pi_{4^{m}})]$.

\subsection{Differential symbolic entropy}
Entropy is a classical approach in quantification of the complexity and is preferable in characterizing real-world time series \cite{Xiong2017}. Differential symbolic entropy, the central concept in our paper, is defined as Shannon entropy of all words' probabilistic distributions as Eq.~(\ref{eq3}), and its normalized form is $h(m)=H(m)/log4^{m}$.
\begin{eqnarray}
\label{eq3}
H(m)=-\sum p(\pi_{i})logp(\pi_{i}), \  where \  p(\pi_{i})\neq 0
\end{eqnarray}

\section{Differential symbolic entropy in chaotic model test}
Logistic map is employed to investigate chaotic detections of DSEn. The two-degree polynomial map, mathematically written as $x_{i+1}=r\cdot x_{i}(1-x_{i})$ , is often cited in chaotic, nonlinear dynamical analysis and used to calculate the properties of random process \cite{may1976}.

\begin{figure}[htb]
  \centering
  \includegraphics[width=8cm,height=4cm]{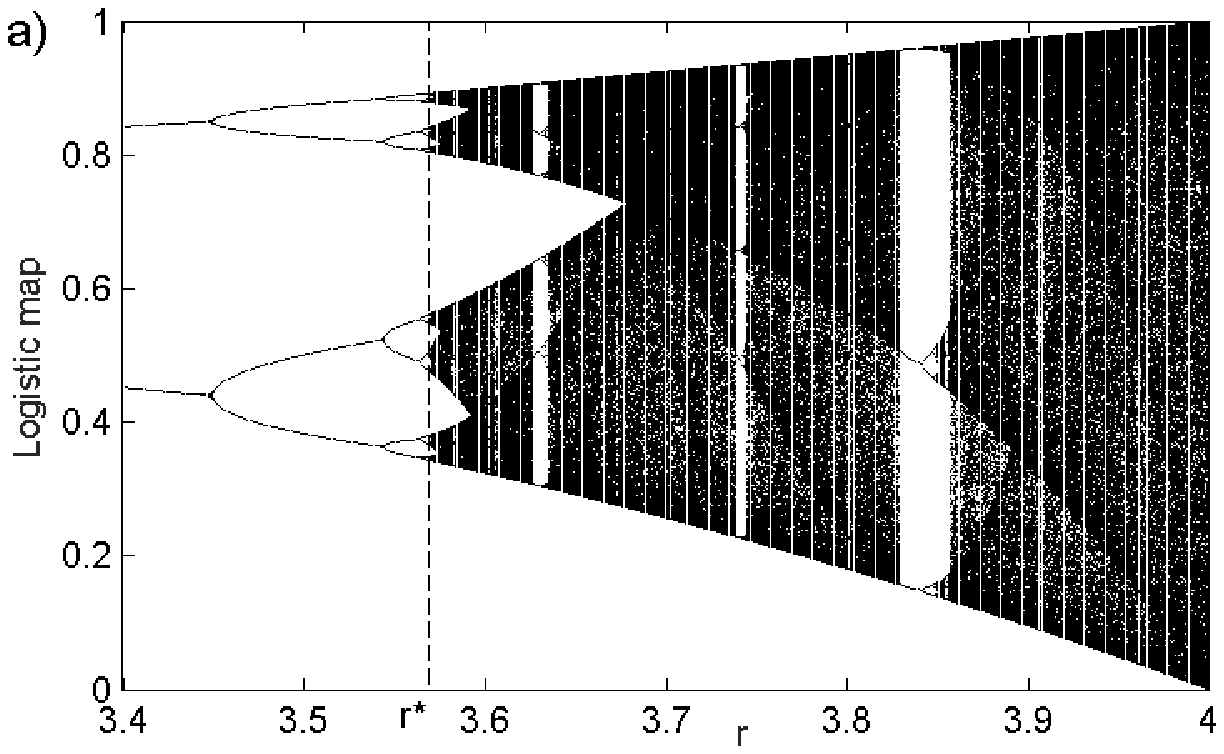}
  \includegraphics[width=8cm,height=4cm]{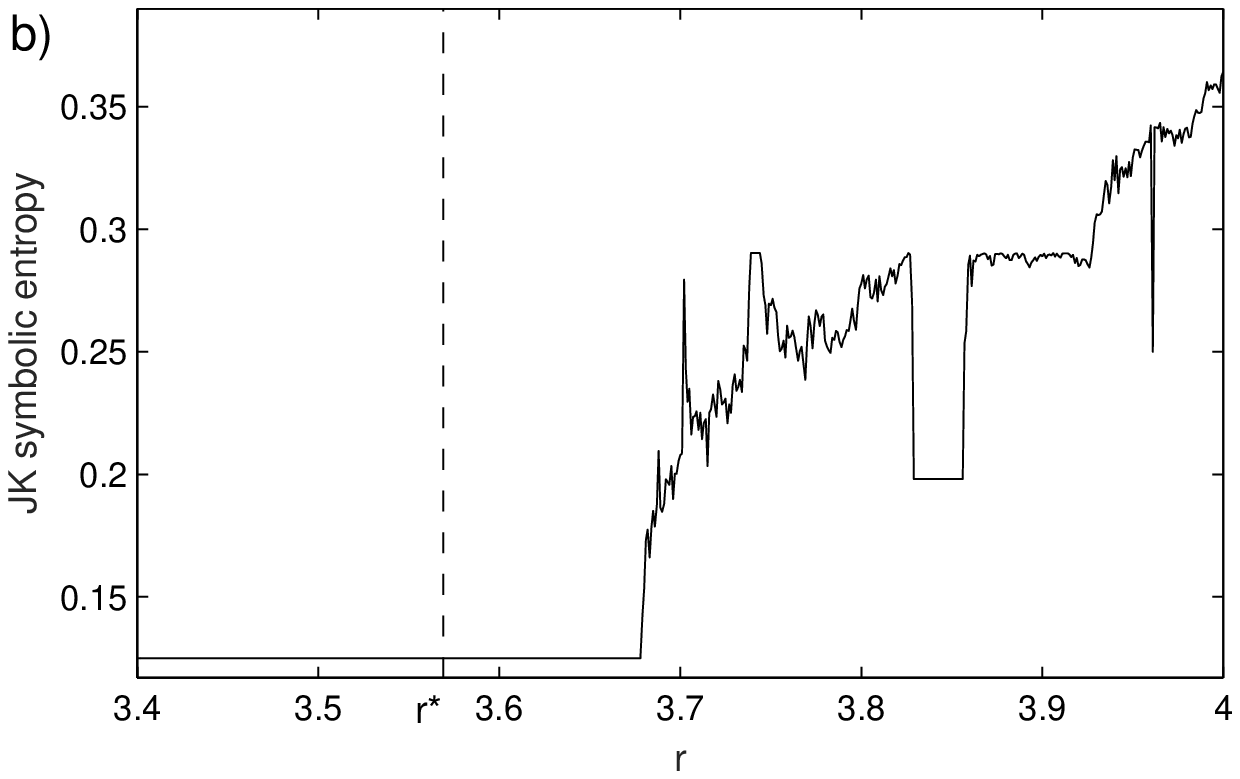}
  \includegraphics[width=8cm,height=4cm]{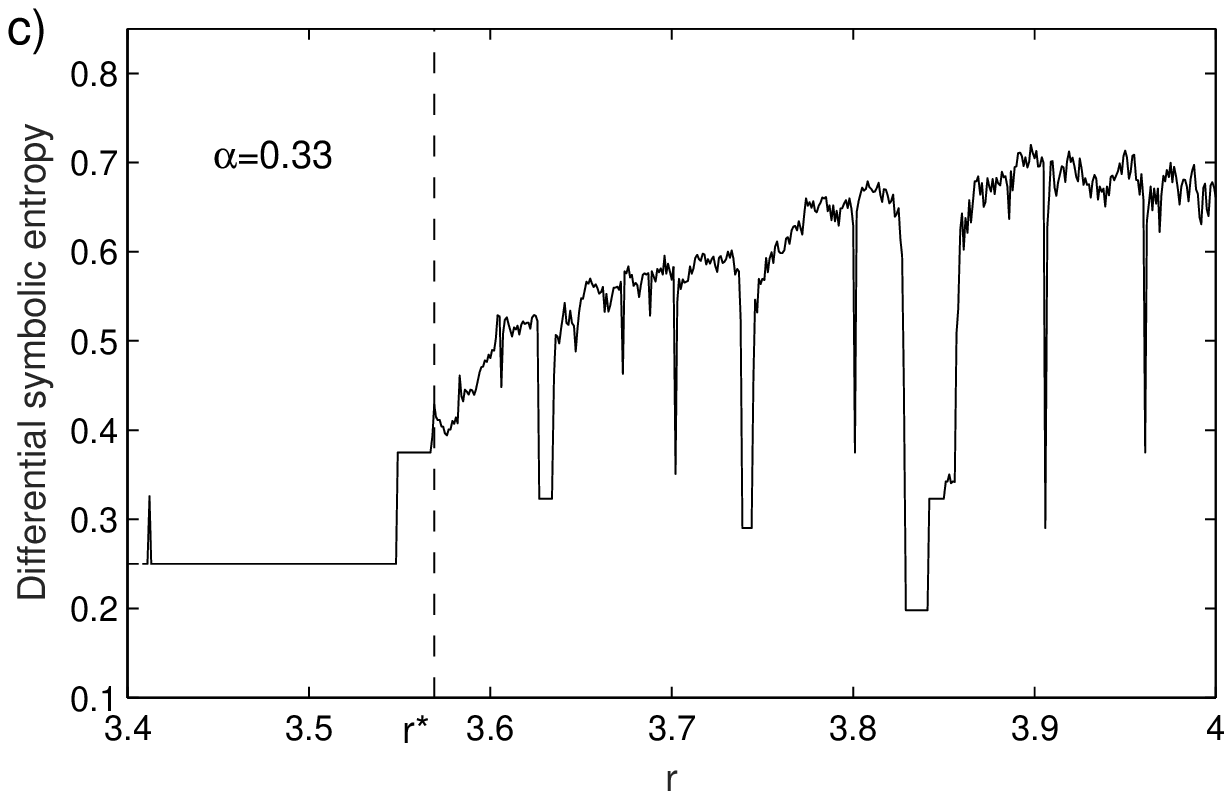}
  \includegraphics[width=8cm,height=4cm]{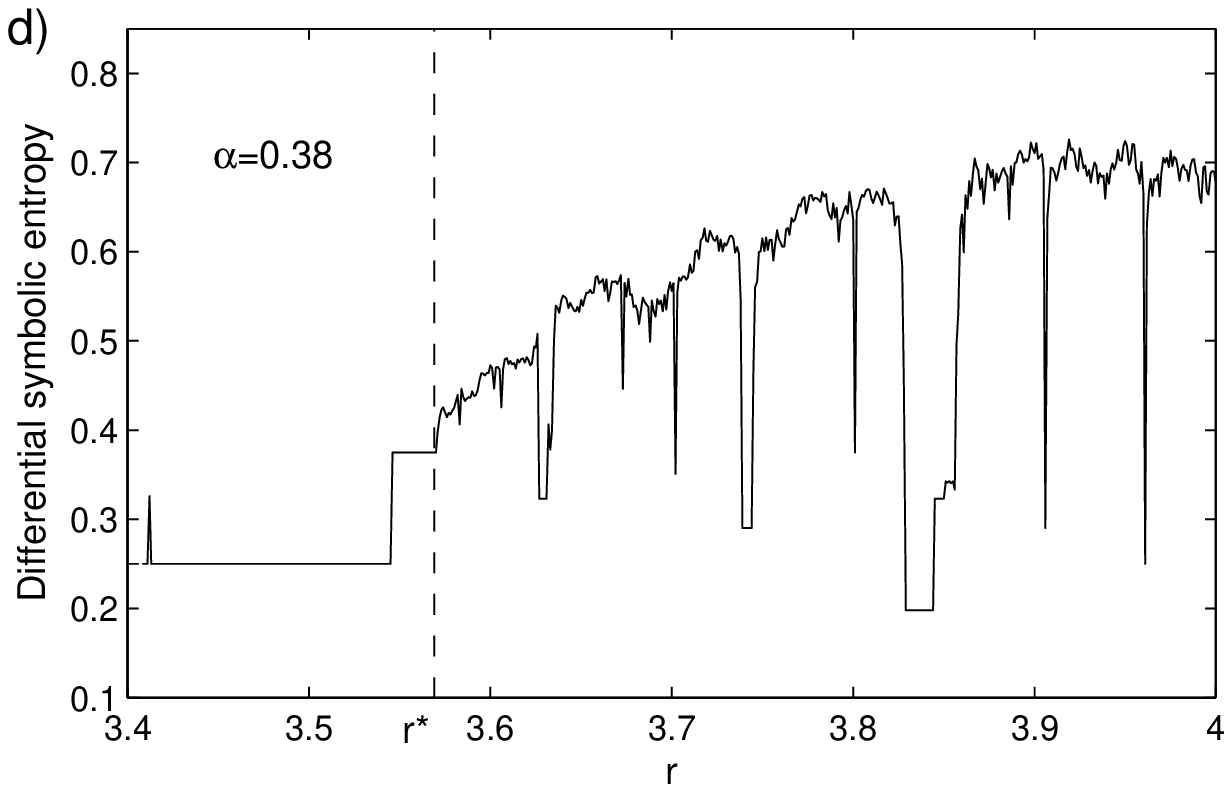}
  \caption{Logistic equations for varying controlling parameter $r$ from 3.4 to 4 with step of $10^{-3}$. Logical map is generated with initial value of 0.03 and length of each sequence is 1200 (the initial value and length of logistic series have no significant influence to chaotic detections of DSEn). As the cut-off point of whether sequence becomes chaotic or not, $r*$ = 3.5669 is marked in each subplot. (a) Bifurcation diagram. (b) JK symbolic entropy. (c) and (d) Differential symbolic entropy with $\alpha$ =0.33 and 0.38.}
  \label{fig2}
\end{figure}

Fig.~\ref{fig2}(a) shows logistic map for varying $r$. In nonlinear complexity detections, DSEn with $\alpha$ between 0.33 and 0.38 is proved to have satisfied chaotic detection at the beginning stage of $r*$ and has increasing entropy with the increasing $r$, showing in subfigures Fig.~\ref{fig2}(c) and (d).

As r becomes to $r*$, JKSEn (JK symbolic entropy), however, shows no change in the early stage as can be seen in Fig.~\ref{fig2}(b). JK symbolic entropy has its starting at r=3.679 which is much bigger than $r*$, so they do not achieve identification of nonlinear features at the very beginning of chaotic logistic series. Moreover, with increasing chaotic features of logistic map, JKSEn have unstable nonlinear dynamics analysis. In $\alpha$=3.702 and between 3.739 and 3.744, JKSEn have abnormally high entropy values, and when $\alpha$ is between 3.857 and 3.927 JKSEn does not increase with enhancing chaotic behaviors of logistic series.

\section{Differential symbolic entropy in heartbeats}
In this section, we test DSEn in nonlinear dynamic complexity extraction of cardiac electric activities. Heart rate, mainly represented RR intervals derived from ECG, is highly irregular and non-stationary \cite{Henriques2016,Costa2008C,Costa2005BM} and contains nonlinear physiological information of cardiac regulation which contributes to scientific researches and clinical applications \cite{Malik1996,Acharya2006}.

Heartbeat intervals of three groups of subjects from Physionet Database \cite{Goldberger2000} are applied to test our differential symbolic entropy. The three kinds of heartbeats, often being applied to test nonlinear approaches \cite{costa2002MP,Li2006,Lo2015,Yao2017D,Costa2005BM}, are collected from patients with severe congestive heart failure \cite{Baim1986} and two groups of healthy young and elderly subjects \cite{Iyengar1996}.

\subsection{Controlling parameter of differential symbolic entropy}
\begin{figure}[htp]
  \centering
  \includegraphics[width=9cm,height=6cm]{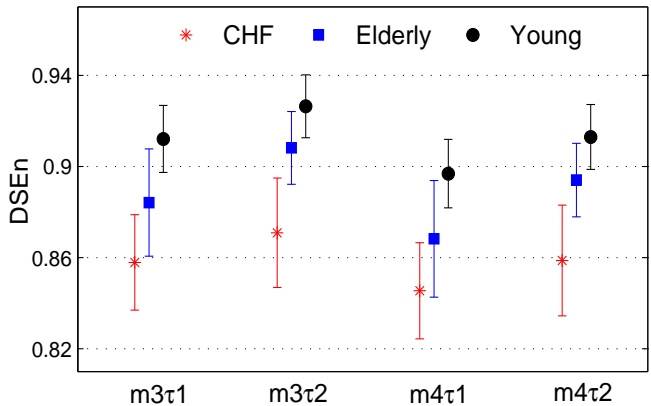}
  \caption{Differential symbolic entropy of three kinds of heart rates. '$m$3$\tau$1' denotes coding length $m$ is 3 and delay $\tau$ is 1.}
  \label{fig3}
\end{figure}

Controlling parameter influences symbolic transformation and nonlinear complexity detections of DSEn in heartbeats. When $\alpha$ is bigger than 0.3, the differential symbolic entropies of three types of heart signals appear Young $>$ Elderly $>$ CHF and remain unchanged. Among all the controlling parameters, $\alpha$ between 0.59 and 0.63 for DSEn have optimum discriminations of the different heartbeats. Differential symbolic entropy of the three kinds of heart rates are shown in Fig.~\ref{fig3}, statistical tests for which are listed in table~\ref{tab1}.

\begin{table}[htb]
\centering
\caption{Statistical tests of DSEn of three kinds of heart rates. P values of DSEn of CHF and healthy young heartbeats, not listed in the table, are less than $2.0*10^{-8}$.}
\label{tab1}
\begin{tabular}{ccc}
\hline
   &$\tau$=1	        &$\tau$=2\\
\hline
CHF-Elderly(m=3) 	&$3.2*10^{-3}$	&$1.5*10^{-5}$ \\
Elderly-Young(m=3)	&$2.0*10^{-4}$	&$1.0*10^{-3}$ \\
\hline
CHF-Elderly(m=4) 	&$1.3*10^{-2}$	&$3.6*10^{-5}$\\
Elderly-Young(m=4)	&$3.2*10^{-4}$  &$8.5*10^{-4}$\\
\hline
\end{tabular}
\end{table}

The basic dynamic complexity relationships of the three kinds of heartbeats characterized by DSEn, healthy young subjects $>$ healthy elderly people $>$ CHF patients, are consistent with the well-accepted complex losing theory that with aging and heart diseases, cardiac modulation associated with pathological alterations that regulate heartbeats will decrease, leading to a loss of nonlinear complexity in heart rates. Reasons for CHF patients' lowest complexity are that CHF damages the patient's heart control systems which leads to complexities fluctuations patterns of heartbeat intervals in CHF patients become quite regular and decreases cardiac inherent nonlinear irregularities. And with increasing age, cardiac functions of the elderly decrease, resulting in loss of dynamic complexity information, so the elderly persons have lower entropy values than the young ones \cite{Costa2005MPE,Costa2008C,Costa2005BM,Buchman2002,Goldberger2002}.

\begin{figure}[htb]
  \centering
  \includegraphics[width=7cm,height=5cm]{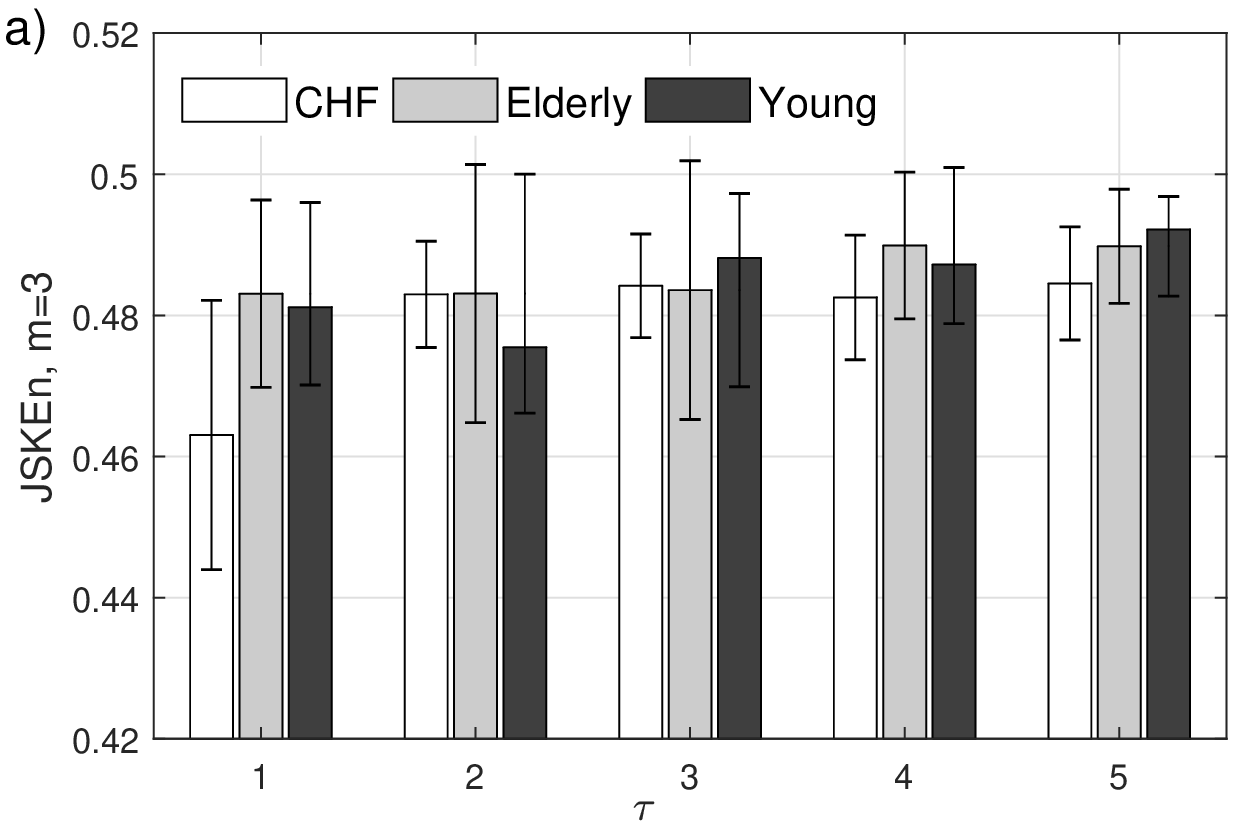}
  \includegraphics[width=7cm,height=5cm]{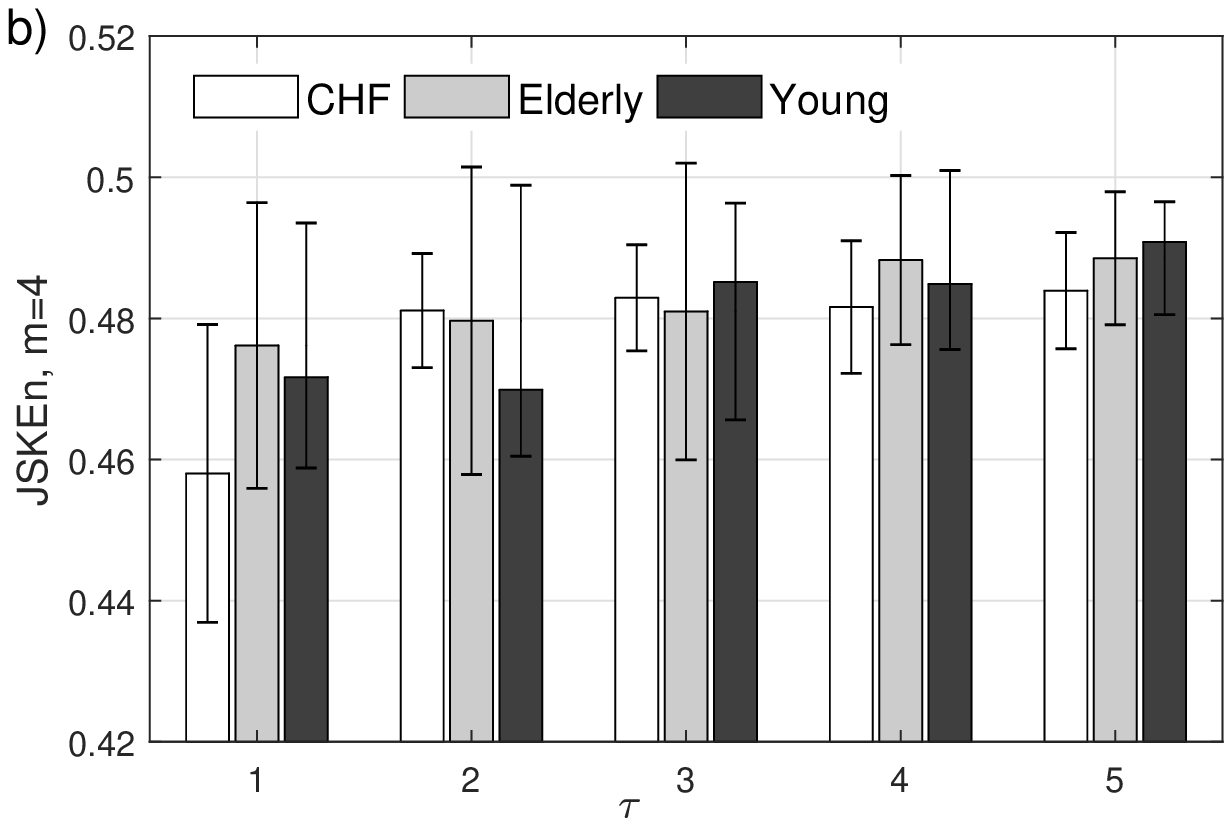}
  \caption{JK symbolic entropy of three kinds of heart rates. Coding lengths are m=3 and 4, coding delay is from 1 to 5. Only when $\tau$=5, JKSEn has similar nonlinearity detections with complex-losing theory.}
  \label{fig4}
\end{figure}

For comparison, JK symbolic entropy of the three kinds of heartbeats is shown in Fig.~\ref{fig4}.

Coding process, particular delay, has great impact on nonlinear analysis of JKSEn in heartbeats analysis. When delay is from 1 to 5, JKSEn of the three kinds of heart rates change too much and have no clear regularity. When delay is 2, JKSEn of the heart signals, healthy young $<$ healthy elderly $<$ CHF, is contrary to complex-losing knowledge about aging and heart diseases \cite{Costa2005MPE,Costa2008C,Costa2005BM,Buchman2002,Goldberger2002}. Only when tau=5, the results are consistent to recent researches, however, the discriminations between CHF and elderly (when m=3 and 4, p=0.065 and 0.142), and between elderly and young people (when m=3 and 4, p=0.32 and 0.408) are not acceptable in statistics.

Compared with JKSEn, DSEn significantly discriminates the three kinds of heartbeats in statistics and its results are consistent with the well-accepted complex losing theory. Two reasons may account for the advantages of DSEn. On one hand, JKSEn only take two elements' relationship into consideration while DSEn applies relationships among three neighboring values which can apply more dynamic information to detect complexities of time series. On the other hand, parameter in the JKSEn is fixed to 1.5 while that in DSEn can be adjusted accordingly. Nonlinear complex processes have different structural or dynamical properties, and different symbolic time series analysis or complexity measures target on different aspects of nonlinear complex systems. One cannot find an optimum controlling parameter for all processes, the disadvantage of fixed range-partitioning symbolization lies in that it can not be adjusted according to different applications.

\subsection{Data length for differential entropy}
In nonlinear complexity detections of chaotic model, the length of logistic sequence does not affect the analysis of DSEn, and in related symbolic entropy researches \cite{Bandt2002,Li2006,Yao2017P}, the nonlinear measures usually do not have high demand for data length and have satisfied complexity extraction in very short time series. To test the influence of date length on the dynamic analysis method, we set data length from 100 to 1800 with step size of 100.

\begin{figure}[htb]
  \centering
  \includegraphics[width=8cm,height=4cm]{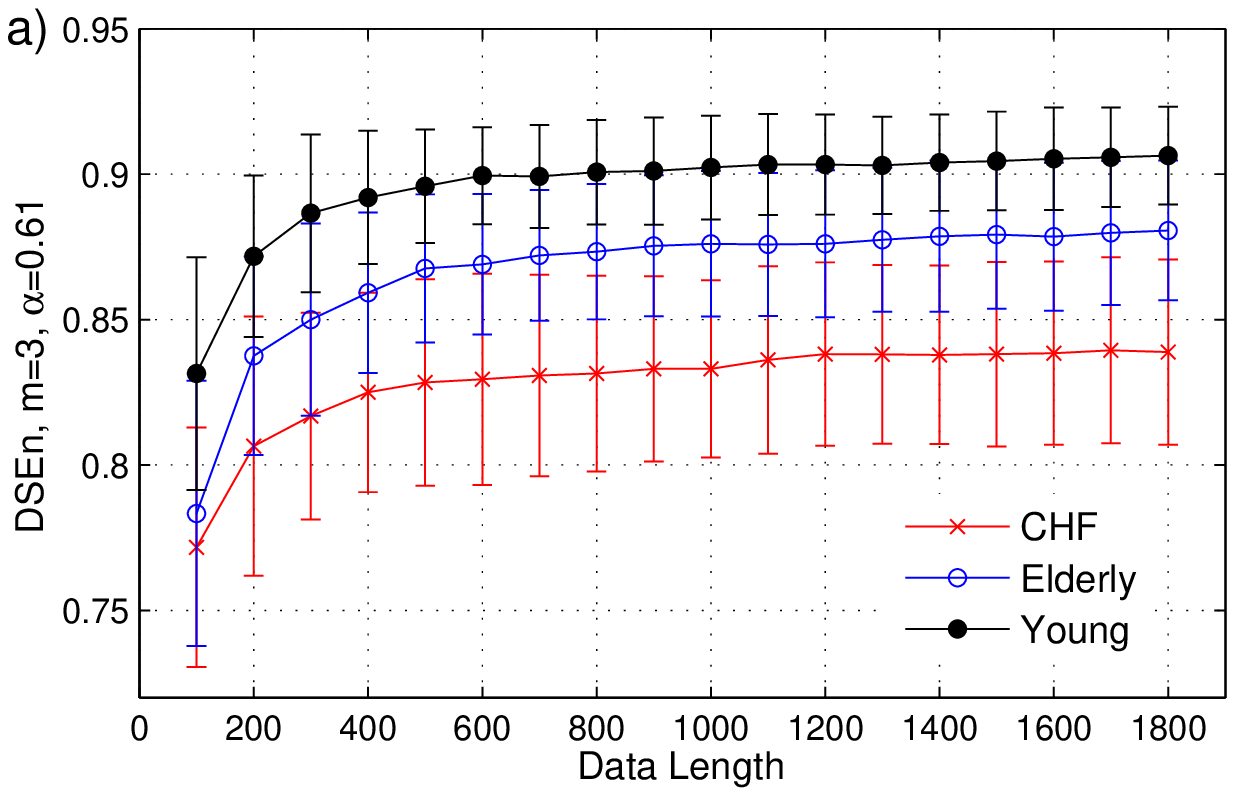}
  \includegraphics[width=8cm,height=4cm]{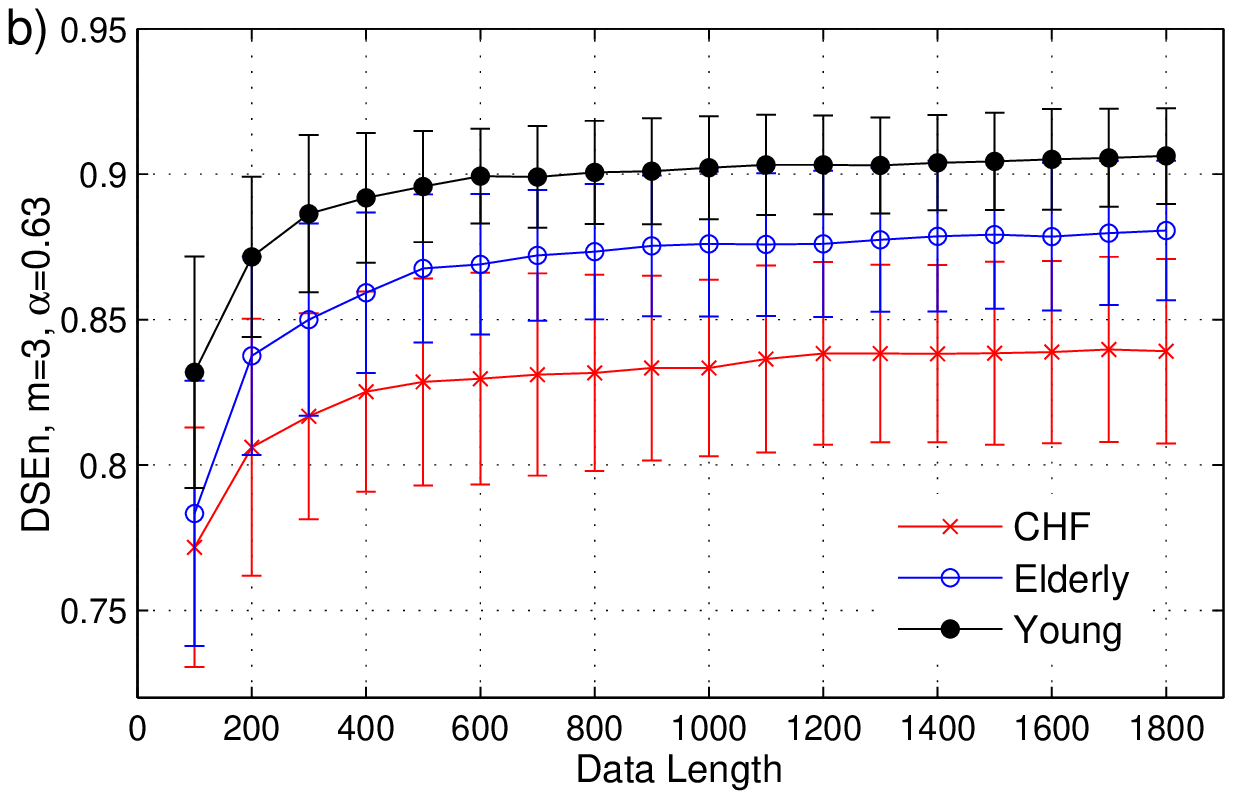}
  \includegraphics[width=8cm,height=4cm]{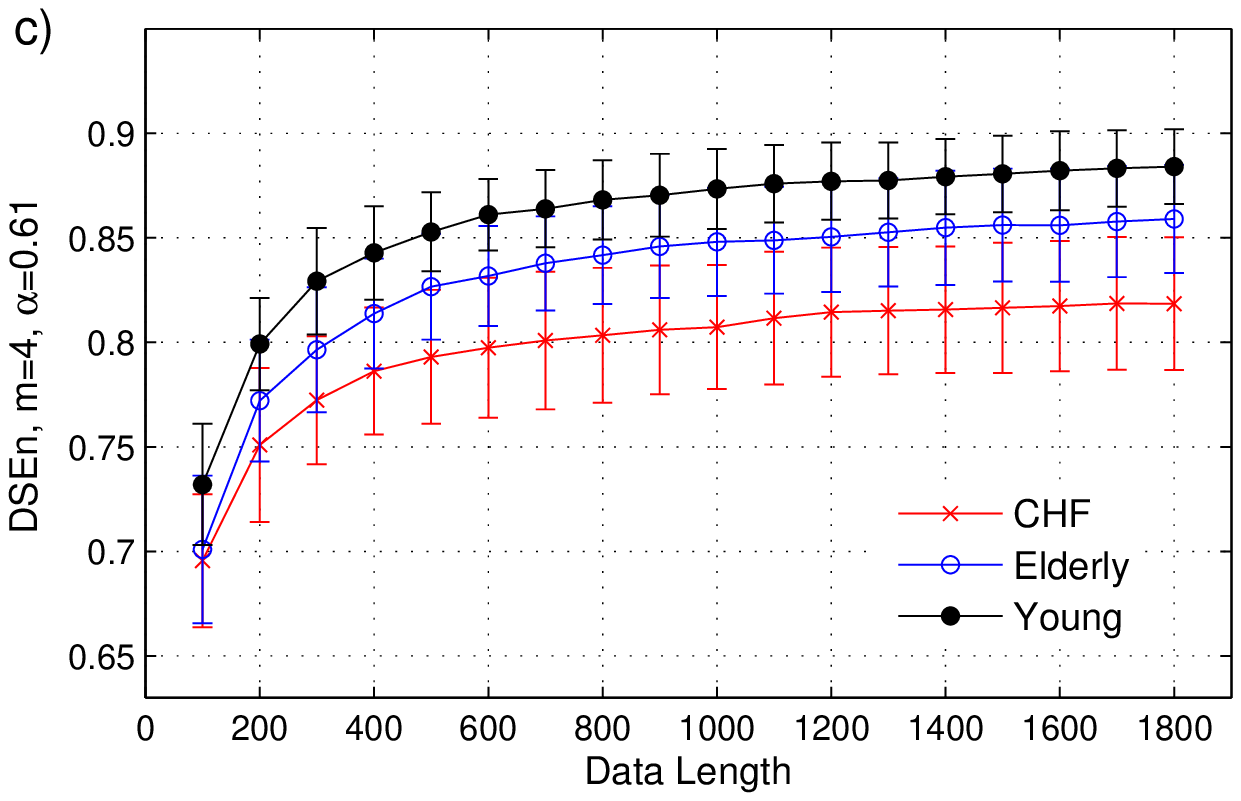}
  \includegraphics[width=8cm,height=4cm]{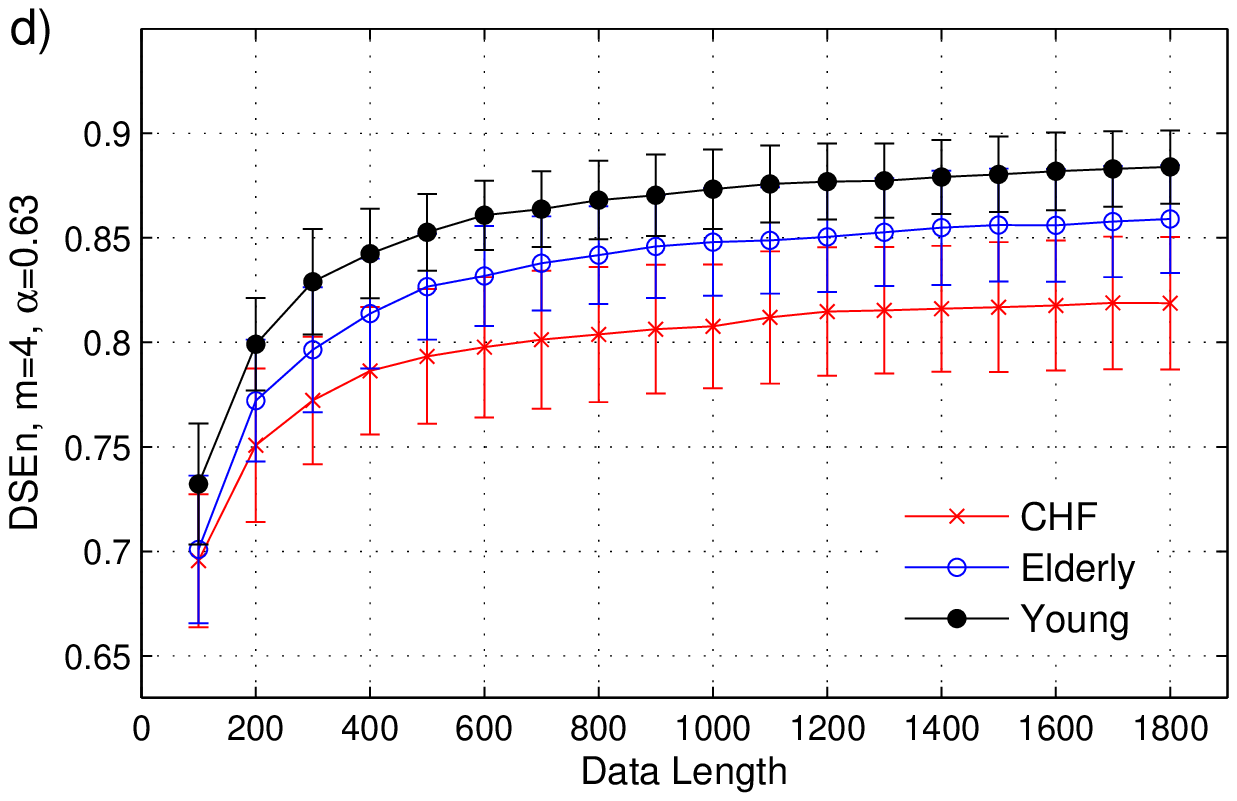}
  \caption{Differential entropy of three types of heartbeats for increasing data length. Encoding lengths of DSEn is 3 and 4 and $\tau=1$. $ \alpha $ of 0.61 and 0.63 are selected referring their satisfied analysis in previous subsection.}
  \label{fig5}
\end{figure}

When data length is only 200, showing in Fig.~\ref{fig5}, three heart rates' nonlinear complexity increases clearly and have clear distinctions among the three kinds of heart rates, the healthy young $ > $ healthy elderly $ > $ CHF patients. When data length increases from 500 to 600, three different heart signals' entropy values have slight growth and gradually tend to be stable and convergent.

\begin{table}[htb]
\centering
\caption{Statistical tests of DSEn of three kinds of heart rates for different data lengths, where '0.000' should be read as 'smaller than 0.001'. P values between DSEn of CHF and healthy young heartbeats are usually less than 0.00001.}
\label{tab2}
\begin{tabular}{ccccc cccc}
\hline
\multicolumn{2}{c}{Data length} 	        &200	        &300	        &400	&500	&600	&700	&800  \\
\hline
\multirow{2}{*}{m=3} &CHF-Elderly 	&0.039	&0.015	&0.006	&0.002	&0.001	&0.001  &0.001\\
                     &Elderly-Young	&0.004	&0.002	&0.001	&0.001	&0.000	&0.001	&0.001 \\
\hline
\multirow{2}{*}{m=4} &CHF-Elderly 	&\textbf{0.092}	&0.040	&0.049	&0.015	&0.004	&0.001	&0.001 \\
                     &Elderly-Young	&0.005	        &0.002	&0.002	&0.002	&0.000	&0.001	&0.001 \\
\hline
\end{tabular}
\end{table}

As indicated by table~\ref{tab2}, the healthy young and elderly subjects, as well as the healthy young and the CHF have been significantly discriminated statistically when data length is only 200. And when data length is 300 or bigger, DSEn of the three groups of heartbeats are significantly different from each other in statistics. In our heartbeats nonlinear complexity analysis, differential symbolic entropy does not have high demands on data length and distinguish three different signals at short data sets. As proved by t tests, DSEn has satisfied nonlinear complexity analysis when data is short to 300 and is not influenced by data length significantly.

From the above analysis, differential symbolic entropy, extracting local dynamic information and detecting nonlinear complexity, has successful chaotic detections in logistic map and real-world physiological heart rates. Moreover, it does not have high requirement to data length and enables short data sets nonlinear dynamic analysis.

Based on researches on chaotic model and real-world physiological signals, we learn that the controlling parameter plays an important role in differential symbolic entropy analysis and it should be adjusted accordingly. In chaotic detections of logistic map, controlling parameters interval [0.33, 0.38] are chosen for their preferable complexity extraction while in nonlinear dynamic analysis of three kinds of heartbeats, controlling parameters having satisfied distinctions should be selected in [0.59, 0.63] according to statistical tests. The controlling parameter, therefore, need to be chosen according to different signals, which at the same time is the drawback of JKSEn. The reason account for variable controlling parameter choices, we suppose, is due to differences in nonlinear structural or dynamical properties of different chaotic models and physiological signals and we can not find a optimal controlling parameter for all different signals.

\section{Conclusions}
Differential symbolic entropy is a nonlinear complexity measure making use of difference-based dynamics in three adjacent elements. The symbolic nonlinear complexity measure with adjustable controlling parameter has satisfied nonlinear analysis in chaotic model and real-world physiological signals and has features of fast and simplicity even for very short data sets.

The complex losing theory of decreased nonlinear complexity in aging and diseased hear rates is validated by DSEn. And our findings suggest that for different structural or dynamical information in complexity systems, controlling parameter in differential symbolic transformation should be adjusted accordingly to extract nonlinear symbolic dynamics.

\section{Acknowledgments}
The work is supported by Project supported by the National Natural Science Foundation of China
(Grant Nos. 61271082, 61401518, 81201161), Jiangsu Provincial Key R \& D Program (Social Development) (Grant No.BE2015700), the Natural Science Foundation of Jiangsu Province (Grant No. BK20141432), Natural Science Research Major Programmer in Universities of Jiangsu Province (Grant No.16KJA310002), Postgraduate Research \& Practice Innovation Program of Jiangsu Province.

\section*{References}

\bibliography{mybibfile}

\begin{thebibliography}{10}
\expandafter\ifx\csname url\endcsname\relax
  \def\url#1{\texttt{#1}}\fi
\expandafter\ifx\csname urlprefix\endcsname\relax\def\urlprefix{URL }\fi
\expandafter\ifx\csname href\endcsname\relax
  \def\href#1#2{#2} \def\path#1{#1}\fi

\bibitem{Schreiber2000}
T.~Schreiber, A.~Schmitz, Surrogate time series, Physica D Nonlinear Phenomena
  142~(3–4) (2000) 346--382.

\bibitem{Storgatz2000}
S.~H. Storgatz, Nonlinear dynamics and chaos : with applications to physics,
  biology, chemistry, and engineering, Perseus Books Publishing, 2000.

\bibitem{Pincus1991}
S.~M. Pincus, Approximate entropy as a measure of system complexity,
  Proceedings of the National Academy of Sciences of the United States of
  America 88~(6) (1991) 2297--2301.

\bibitem{Richman2000}
J.~S. Richman, J.~R. Moorman, Physiological time-series analysis using
  approximate entropy and sample entropy, American Journal of Physiology Heart
  \& Circulatory Physiology 278~(6) (2000) 2039--2049.

\bibitem{Keller2009}
K.~Keller, M.~Sinn, Kolmogorov sinai entropy from the ordinal viewpoint,
  Physica D Nonlinear Phenomena 239~(12) (2009) 997--1000.

\bibitem{Chen2009}
W.~Chen, J.~Zhuang, W.~Yu, Z.~Wang, Measuring complexity using fuzzyen, apen,
  and sampen, Medical Engineering \& Physics 31~(1) (2009) 61--68.

\bibitem{Bandt2002}
C.~Bandt, B.~Pompe, Permutation entropy: a natural complexity measure for time
  series, Physical Review Letters 88~(17) (2002) 174102.

\bibitem{Bandt2016}
C.~Bandt, Permutation Entropy and Order Patterns in Long Time Series, Springer,
  2016, pp. 61--73.

\bibitem{costa2002MP}
M.~D. Costa, A.~L. Goldberger, C.~K. Peng, Multiscale entropy analysis of
  complex physiologic time series, Physical Review Letters 89~(6) (2002)
  068102.

\bibitem{Costa2005MPE}
M.~Costa, A.~L. Goldberger, C.~K. Peng, Multiscale entropy analysis of
  biological signals, Physical Review E 71~(2 Pt 1) (2005) 021906.

\bibitem{Li2015}
P.~Li, C.~Liu, K.~Li, D.~Zheng, C.~Liu, Y.~Hou, Assessing the complexity of
  short-term heartbeat interval series by distribution entropy, Medical \&
  Biological Engineering \& Computing 53~(1) (2015) 77--87.

\bibitem{Isler2007}
Y.~Isler, M.~Kuntalp, Combining classical hrv indices with wavelet entropy
  measures improves to performance in diagnosing congestive heart failure,
  Computers in Biology and Medicine 37~(10) (2007) 1502--1510.

\bibitem{Yao2014}
W.~Yao, T.~Liu, J.~Dai, J.~Wang, Multiscale permutation entropy analysis of
  electroencephalogram, Acta Phys. Sin. 63~(7) (2014) 78704.

\bibitem{Li2006}
J.~Li, X.~Ning, Dynamical complexity detection in short-term physiological
  series using base-scale entropy, Physical Review E 73~(5 Pt 1) (2006) 052902.

\bibitem{Hao1991}
B.~L. Hao, Symbolic dynamics and characterization of complexity, Physica D
  Nonlinear Phenomena 51~(1–3) (1991) 161--176.

\bibitem{Daw2003}
C.~S. Daw, C.~E.~A. Finney, E.~R. Tracy, A review of symbolic analysis of
  experimental data, Review of Scientific Instruments 74~(2) (2003) 915--930.

\bibitem{Staniek2008}
M.~Staniek, K.~Lehnertz, Symbolic transfer entropy, Physical Review Letters
  100~(15) (2008) 158101.

\bibitem{Yao2017P}
Y.~Wenpo, W.~Jun, Multi-scale symbolic transfer entropy analysis of eeg,
  Physica A: Statistical Mechanics and its Applications 484 (2017) 276--281.

\bibitem{Lo2015}
M.~T. Lo, Y.~C. Chang, C.~Lin, H.~W. Young, Y.~H. Lin, Y.~L. Ho, C.~K. Peng,
  K.~Hu, Outlier-resilient complexity analysis of heartbeat dynamics,
  Scientific Reports 5 (2015) 8836.

\bibitem{Yao2017D}
W.~Yao, J.~Wang, Double symbolic joint entropy in nonlinear dynamic complexity
  analysis, Aip Advances 7~(7) (2017) 075313.

\bibitem{Daw1998}
C.~S. Daw, C.~E.~A. Finney, K.~Nguyen, J.~S. Halow, Symbol statistics: A new
  tool for understanding multiphase flow phenomena, International Mechanical
  Engineering Congress \& Exposition 34~(2) (1998) 299--315.

\bibitem{Finney1998}
C.~E.~A. Finney, K.~Nguyen, C.~S. Daw, J.~S. Halow, Symbol-sequence statistics
  for monitoring fluidization, International Mechanical Engineering Congress \&
  Exposition 83~(7) (1998) 687--704.

\bibitem{Zanin2012}
M.~Zanin, L.~Zunino, O.~A. Rosso, D.~Papo, Permutation entropy and its main
  biomedical and econophysics applications: A review, Entropy 14~(8) (2012)
  1553--1577.

\bibitem{Wessel2000}
N.~Wessel, C.~Ziehmann, J.~Kurths, U.~Meyerfeldt, A.~Schirdewan, A.~Voss,
  Short-term forecasting of life-threatening cardiac arrhythmias based on
  symbolic dynamics and finite-time growth rates, Physical Review E 61~(1)
  (2000) 733--739.

\bibitem{Wessel2007}
N.~Wessel, H.~Malberg, R.~Bauernschmitt, J.~Kurths, Nonlinear methods of
  cardiovascular physics and their clinical applicability, International
  Journal of Bifurcation \& Chaos 17~(10) (2007) 3325--3371.

\bibitem{Kurths1995}
J.~Kurths, A.~Voss, P.~Saparin, A.~Witt, H.~J. Kleiner, N.~Wessel, Quantitative
  analysis of heart rate variability, Chaos 5~(1) (1995) 88--94.

\bibitem{Xiong2017}
W.~Xiong, L.~Faes, P.~C. Ivanov, Entropy measures, entropy estimators, and
  their performance in quantifying complex dynamics: Effects of artifacts,
  nonstationarity, and long-range correlations, Physical Review E 95~(6) (2017)
  062114.

\bibitem{may1976}
R.~M. May, Simple mathematical models with very complicated dynamics, Nature
  261~(5560) (1976) 459--467.

\bibitem{Henriques2016}
T.~S. Henriques, S.~Mariani, A.~Burykin, F.~Rodrigues, T.~F. Silva, A.~L.
  Goldberger, Multiscale poincaré plots for visualizing the structure of
  heartbeat time series, BMC Medical Informatics and Decision Making 16~(1)
  (2016) 1--17.

\bibitem{Costa2008C}
M.~D. Costa, C.~K. Peng, A.~L. Goldberger, Multiscale analysis of heart rate
  dynamics: Entropy and time irreversibility measures, Cardiovascular
  Engineering 8~(2) (2008) 88--93.

\bibitem{Costa2005BM}
M.~Costa, A.~L. Goldberger, C.-K. Peng, Broken asymmetry of the human
  heartbeat: loss of time irreversibility in aging and disease, Physical Review
  Letters 95~(19) (2005) 198102.

\bibitem{Malik1996}
M.~Malik, A.~J. Camm, J.~T. Bigger, G.~Breithardt, S.~Cerutti, R.~J. Cohen,
  P.~Coumel, E.~L. Fallen, H.~L. Kennedy, R.~E. Kleiger, F.~Lombardi,
  A.~Malliani, A.~J. Moss, J.~N. Rottman, G.~Schmidt, P.~J. Schwartz, D.~H.
  Singer, Heart rate variability: standards of measurement, physiological
  interpretation and clinical use, Circulation 93~(5) (1996) 1043--1065.

\bibitem{Acharya2006}
U.~R. Acharya, K.~P. Joseph, N.~Kannathal, C.~M. Lim, J.~S. Suri, Heart rate
  variability: a review, Medical \& Biological Engineering \& Computing 44~(12)
  (2006) 1031--1051.

\bibitem{Goldberger2000}
A.~L. Goldberger, L.~A. Amaral, L.~Glass, J.~M. Hausdorff, P.~C. Ivanov, R.~G.
  Mark, J.~E. Mietus, G.~B. Moody, C.~K. Peng, H.~E. Stanley, Physiobank,
  physiotoolkit, and physionet: components of a new research resource for
  complex physiologic signals, Circulation 101~(23) (2000) 215--220.

\bibitem{Baim1986}
D.~S. Baim, W.~S. Colucci, E.~S. Monrad, H.~S. Smith, R.~F. Wright, A.~Lanoue,
  D.~F. Gauthier, B.~J. Ransil, W.~Grossman, E.~Braunwald, Survival of patients
  with severe congestive heart failure treated with oral milrinone, Journal of
  the American College of Cardiology 7~(3) (1986) 661--670.

\bibitem{Iyengar1996}
N.~Iyengar, C.~K. Peng, R.~Morin, A.~L. Goldberger, L.~A. Lipsitz, Age-related
  alterations in the fractal scaling of cardiac interbeat interval dynamics,
  American Journal of Physiology 271~(2) (1996) 1078--1084.

\bibitem{Buchman2002}
T.~G. Buchman, The community of the self, Nature 420~(6912) (2002) 246.

\bibitem{Goldberger2002}
A.~L. Goldberger, C.~K. Peng, L.~A. Lipsitz, What is physiologic complexity and
  how does it change with aging and disease?, Neurobiology of Aging 23~(1)
  (2002) 23--26.

\end{thebibliography}

\end{document}